\documentclass[%
superscriptaddress,
preprint,
 amsmath,amssymb,
 aps,
prb,
]{revtex4-1}

\usepackage{graphicx}
\usepackage{dcolumn}
\usepackage{bm}
\usepackage{float}

\begin{document}

\title{Commensurate Stripes and Phase Coherence in Manganites Revealed with Cryogenic Scanning Transmission Electron Microscopy}
\author{Ismail El Baggari} 
\affiliation{Department of Physics, Cornell University, Ithaca, NY 14853, USA}%
\author{Benjamin H. Savitzky} 
\affiliation{Department of Physics, Cornell University, Ithaca, NY 14853, USA}%
\author{Alemayehu S. Admasu} 
\affiliation{Rutgers Center for Emergent Materials and Department of Physics and Astronomy, Rutgers University, Piscataway, NJ 08854, USA}%
\author{Jaewook Kim} 
\affiliation{Rutgers Center for Emergent Materials and Department of Physics and Astronomy, Rutgers University, Piscataway, NJ 08854, USA}%
\author{Sang-Wook Cheong} 
\affiliation{Rutgers Center for Emergent Materials and Department of Physics and Astronomy, Rutgers University, Piscataway, NJ 08854, USA}%
\author{Robert Hovden} 
\affiliation{School of Applied and Engineering Physics, Cornell University, Ithaca, NY 14853, USA}%
\affiliation{Current address: Department of Materials Science and Engineering, University of Michigan, Ann Arbor, MI 48109, USA}
\author{Lena F. Kourkoutis} 
\affiliation{School of Applied and Engineering Physics, Cornell University, Ithaca, NY 14853, USA}%
\affiliation{Kavli Institute for Nanoscale Science, Cornell University, Ithaca, NY 14853, USA}

\maketitle



\newpage

\section*{Abstract}

\textbf{
Incommensurate charge order in hole-doped oxides is intertwined with exotic phenomena such as colossal magnetoresistance, high-temperature superconductivity, and electronic nematicity.
Here, we map at atomic resolution the nature of incommensurate order in a manganite using scanning transmission electron microscopy at room temperature and cryogenic temperature ($\sim$ 93K).
In diffraction, the ordering wavevector changes upon cooling, a behavior typically associated with incommensurate order.
However, using real space measurements, we discover that the underlying ordered state is lattice-commensurate at both temperatures.
The cations undergo picometer-scale ($\sim$6-11 pm) transverse displacements, which suggests that charge-lattice coupling is strong and hence favors lattice-locked modulations. 
We further unearth phase inhomogeneity in the periodic lattice displacements at room temperature, and emergent phase coherence at 93K. 
Such local phase variations not only govern the long range correlations of the charge-ordered state, but also results in apparent shifts in the ordering wavevector. 
These atomically-resolved observations underscore the importance of lattice coupling and provide a microscopic explanation for putative "incommensurate" order in hole-doped oxides. 
}

\newpage

\section*{Introduction}

Charge-ordered phases permeate the phase diagrams of strongly correlated systems such as cuprate high-temperature superconductors, colossal magnetoresitive manganites, and 2D transition-metal dichalcogenides.\cite{Chen1996,Uehara1999,Chen1999,Wu2011,Ghiringhelli2012,Chang2012,kusmartseva2009,li2015}. 
Charge order is a modulation of the electron density that breaks lattice translational symmetry and induces periodic lattice displacements via electron-lattice coupling. 
Bulk measurements have unearthed complex interactions between charge order and electronic phases including direct competition with superconductivity or mediation of colossal magnetoresistance, 
which highlights the importance of charge order in understanding and manipulating novel phases of matter\cite{Uehara1999,Chang2012,kusmartseva2009,li2015,Zhang2016}. 

While the precise microscopic mechanism of charge ordering remains under intense scrutiny, measurements of modulation wavevectors in various materials have establishd a tendency towards incommensurate order\cite{Chen1996,Chen1999,Milward2005,Ghiringhelli2012,Feng2015,Comin2016,Mesaros2011,Mesaros2016}.
The presence of incommensuration coincides with the emergence of competing phases such as superconductivity and has motivated interrogation of the role of Fermi surface instabilities\cite{Milward2005,Chuang2001,morosan2006superconductivity,Sipos2008,wise2008charge}. 
Scattering experiments, for instance, have measured changes in the positions of incommensurate wavevectors as a function of temperature, pressure or doping, which is thought to reflect changes in the nesting of the Fermi surface. 
However, coupling to quenched impurities and attendant order parameter fluctuations profoundly alter correlation lengths and symmetry, and complicate experimental interpretation of reciprocal space behavior\cite{Kivelson2003,DelMaestro2006,Mesaros2016}.

Bi$_{1-x}$Sr$_{x-y}$Ca$_{y}$MnO$_{3}$ (BSCMO) is a model charge-ordered manganite with a high, tunable transition temperature (T$_{\mathrm{c}}$)\cite{Kim2007}.
Dark-field transmission electron microscopy has previously visualized striped superstructures in manganites, interpreting contrast as the ordering of holes on alternating manganese sites (Mn$^{3+}$-Mn$^{4+}$)\cite{Chen1996,Chen1999,Milward2005}.
Other experiments based on transport and electron diffraction advance that charge order corresponds to a uniform charge density wave with small valence modulations\cite{Loudon2005,Cox2008}.
The debate is partly motivated by the need to reconcile incommensurate wavevectors with discrete charge ordering.
Understanding striped phases in manganites is further complicated by the ubiquity of quenched disorder and nanoscale phase inhomogeneity.
Optical studies on BSCMO, for instance, suggest that long-range correlations gradually develop below T$_{c}$, an indication that charge order is disturbed by temperature-dependent spatial inhomogeneity\cite{Rubhausen2000}.
To achieve a microscopic understanding of charge-ordered states, atomic-scale characterization of individual degrees of freedom is necessary. 

Room temperature scanning transmission electron microscopy (STEM) enables measurements of atomic column positions with picometer precision\cite{Yankovich2014} and has been used to unravel, for instance, novel ferroelectric behavior in oxides\cite{nelson2011spontaneous,zhang2013direct}.  
Recently, we have revealed periodic lattice displacements associated with charge ordering at room temperature in BSCMO (T$_{c}$ $\sim$ 300K) and visualized nanoscale inhomogeneity in the modulation field\cite{Savitzky2017}.
Here, we demonstrate cryogenic STEM imaging with sub-Angstrom resolution ($\sim$0.78\r{A}) and sufficient signal-to-noise ratio to visualize the charge-ordered state in BSCMO well below T$_{c}$.
Earlier cryogenic STEM studies have observed ordering phenomena \cite{Klie2007,Hovden2016}, however, stage instability, limited resolution, and low signal-to-noise ratio have precluded mapping of picometer-scale lattice behavior. 

In BSCMO, we find that the lattice modulations are commensurate over nanometer scales at both room temperature and cryogenic temperature. 
In contrast, area-averaged diffraction measurements suggest that the modulation wavevector is incommensurate and that it varies with temperature.
By extracting the phase of the lattice modulations, we uncover nanoscale phase variations which not only control long-range ordering, but also result in wavevector shifts in reciprocal space. 
Upon cooling, the phase field becomes more homogeneous and the wavevector converges to the underlying, commensurate value. 
Our observations support that strong lattice coupling favors commensurate order and that local inhomogeneity fundamentally alters macroscopic measurements of ordering wavevectors. 
More generally, cryogenic STEM paves the way for direct visualization of correlated lattice order with picometer precision over atomic and nanometer scales.

\subsection*{Results}
\begin{figure}
  \includegraphics[width=.7\linewidth]{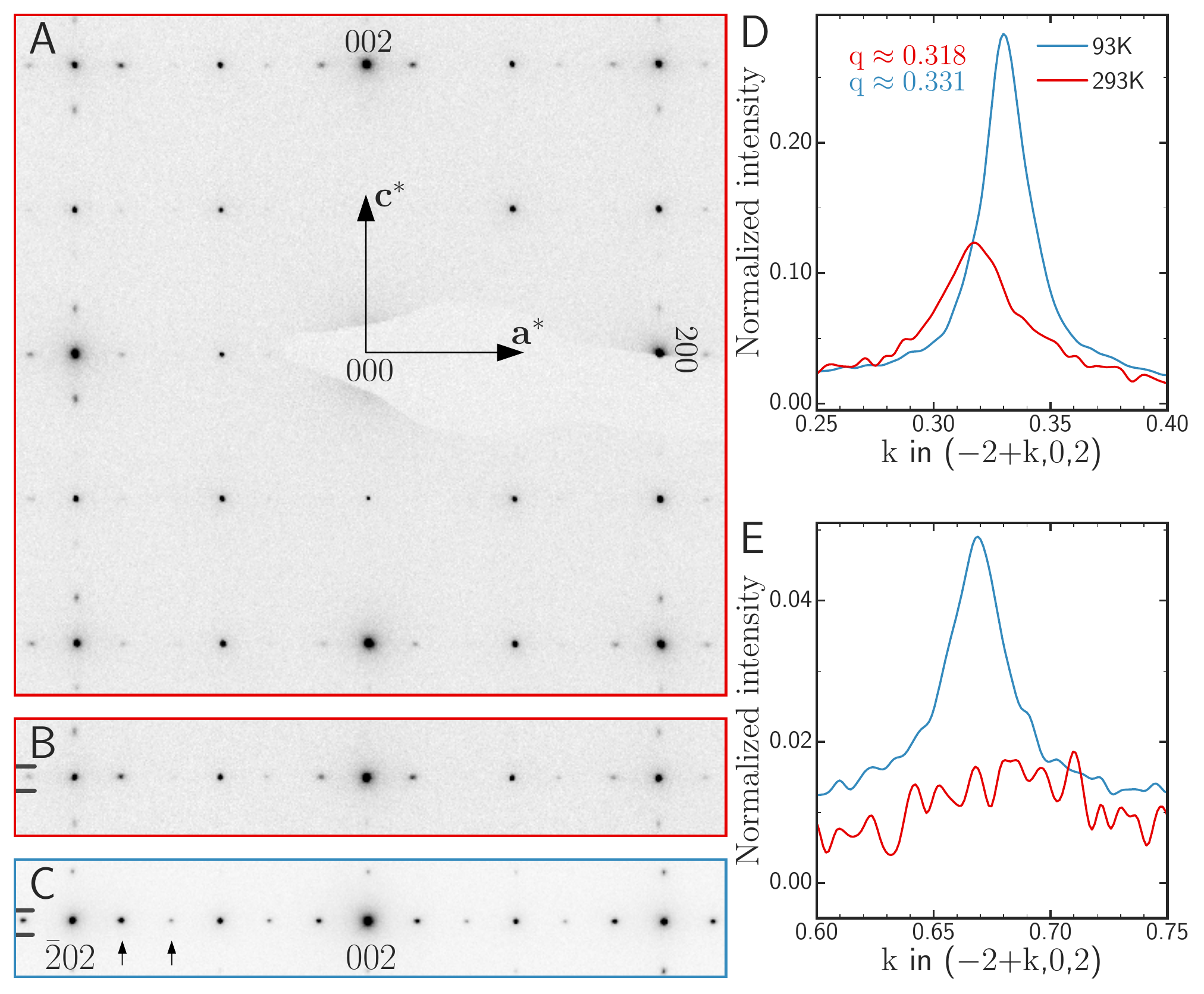}
  \caption{Long range order and wavevector variation upon cooling from 293K to 93K.
  (A) Typical electron diffraction pattern of Bi$_{1-x}$Sr$_{x-y}$Ca$_{y}$MnO$_{3}$ at 293K.
  (B), (C) Section of the diffraction pattern from $\bar{2}02$ to $202$ at 293K and 93K, respectively. 
  (D) Projected intensity of the superlattice peak, $q$, near the $\bar{2}02$ Bragg peak along the $\mathbf{a^{*}}$ direction.
  The intensity is integrated between the tick marks in (B) and (C) and is normalized by the $\bar{2}02$ Bragg peak intensity.
  Upon cooling, there is a change in the wavevector, from $q$= 0.318 reciprocal lattice units (r.l.u) at 293K to $q$=0.331 r.l.u at 93K, a behavior typically associated with incommensurate order. 
  (E) Projected intensity near the $2q$ peak along the $\mathbf{a^{*}}$ direction.
The intensity is integrated between the tick marks in (B) and (C) and is normalized by the $\bar{2}02$ Bragg peak intensity.
  }
  \label{F:Fig1}
\end{figure}

Bi$_{1-x}$Sr$_{x-y}$Ca$_{y}$MnO$_{3}$ (BSCMO) single crystals are grown using the flux method, using Bi$_2$O$_3$, CaCO$_3$, SrCO$_3$, and Mn$_2$O$_3$, as reported previously\cite{Savitzky2017}.
We measure the composition of BSCMO to be approximately $x=0.65$ and $y=0.47$ and find that a resistivity anomaly associated with charge ordering occurs at T$_{c}$ $\sim$ 300K (\textit{SI Appendix}, Fig. S1).
Figure~\ref{F:Fig1}A shows a typical, room temperature (293K) electron diffraction pattern of BSCMO, which we index in the $Pnma$ space group.
In addition to crystalline Bragg peaks, we observe superlattice peaks at $\pm\mathbf{q}$ indicating the presence of a periodic modulation at both 293K (Fig.~\ref{F:Fig1}B) and 93K (Fig.~\ref{F:Fig1}C). 
The projected intensity of the (-2+$q$,0,2) peak along $\mathbf{a^{*}}$ at 293K (red) and 93K (blue) is shown in Fig.~\ref{F:Fig1}D. 
At each temperature, the intensity profile is obtained by integrating between the tick marks in Figs.~\ref{F:Fig1}B and~\ref{F:Fig1}C, and is normalized by the respective integrated intensity of the $\bar{2}02$ Bragg peak.
We note a clear shift in the superlattice peak position upon cooling, from $q=0.318$ reciprocal lattice units (r.l.u) at 293K to $q=0.331$ r.l.u. at 93K (\textit{SI Appendix}, Fig. S7). 
By fitting multiple satellite in the diffraction pattern, we obtain the average wavevector and its uncertainty at each temperature.
The average wavevector is $q=0.314\pm 0.003$ r.l.u at 293K and $q=0.332 \pm 0.001$ r.l.u at 93K.
The magnitude of the wavevector shift is $\delta q = 0.018 \pm 0.003$ r.l.u. 
Temperature-dependent wavevector variations, as observed here, are typically considered an indication of incommensurate order.

In addition to a shift in the wavevector, the superlattice peak also exhibits a clear increase in intensity, $I$, and a decrease in the full-width-at-half maximum, $\sigma_{k}$, at low temperature.
Fitting a Lorentzian function and a linear background to the projected $q$ peak, we find that $I\mathrm{(93K)}$/$I\mathrm{(293K)}\sim 1.46$ and $\sigma_{k}$(93K)/$\sigma_{k}$(293K) $\sim$ 0.57 (\textit{SI Appendix}, Fig. S7).
The weaker $2q$ peak, while almost undetectable at room temperature, is relatively sharp and well defined at 93K (Fig.~\ref{F:Fig1}E). 
These observations indicate that charge order strength and correlations increase well below T$_{c}$.

From previous real space measurements at room temperature, we uncovered periodic lattice displacements (PLD) associated with charge ordering in BSCMO\cite{Savitzky2017}.   
Periodic lattice displacements may be described by the order parameter
\begin{equation*}
\mathbf{\Delta(r)}= \mathfrak{Re}\{\mathbf{A(r)}e^{i\phi(\mathbf{r})}e^{\mathbf{iq.r}}  \} 
\end{equation*}
where $\mathbf{A(r)}$ is the displacement vector, $\mathbf{q}$ is the wavevector, and $\phi$(\textbf{r}) is the phase.
The modulations give rise to complex-valued satellite peaks in the Fourier transform (FT) which are given by $ S(\mathbf{k})\sim \sum\limits_{\{\mathbf{r}\}} \mathrm{exp}[i \mathbf{(k\pm q) . r}] \ \mathrm{exp}[i\phi(\mathbf{r})]$ where $\{\mathbf{r}\}$ is the set of lattice positions (\textit{SI Appendix, Structure Factor of Periodic Lattice Displacements}). 
Phase information not only encodes the particular realization of $\mathbf{\Delta(r)}$, but also its disorder.
Since diffraction experiments probe the intensity ($I(\mathbf{k}) \sim |S(\mathbf{k})|^{2}$), they are insensitive to phase information and provide instead globally-averaged measurements of correlation lengths, intensities, and wavevectors\cite{Chang2012}. 
However, the microscopic picture of incommensurate modulations and their temperature (or doping) dependence may be linked to phase information\cite{Mesaros2016}. 

\begin{figure}[h!]
  \includegraphics[width=.6\linewidth]{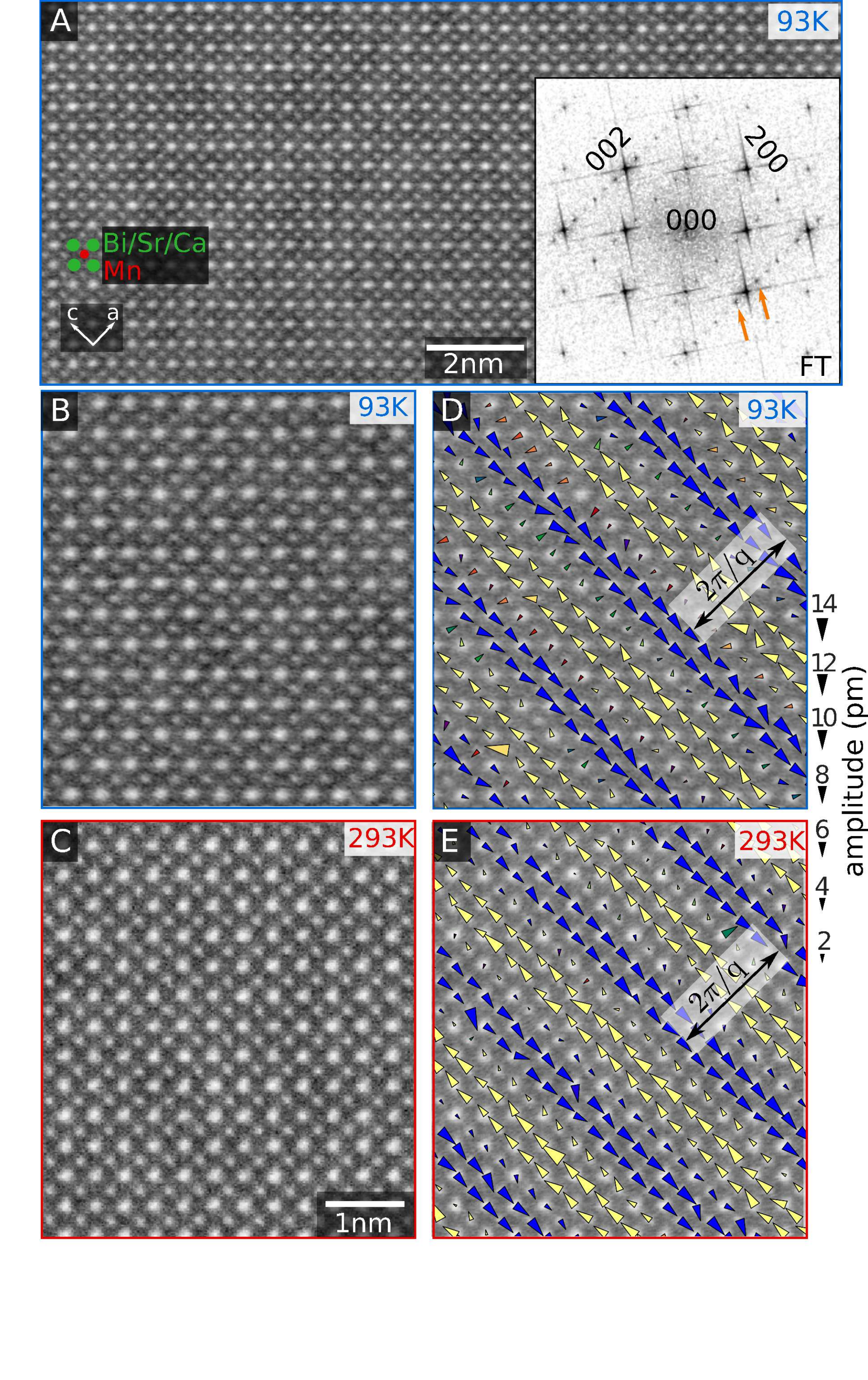}
  \caption{Locally commensurate, picometer-scale periodic lattice displacements at room and cryogenic temperatures. (A) HAADF STEM image and its Fourier transform (inset) at 93K.
  Bi/Sr/Ca columns (green) and Mn columns (red) are clearly resolved. 
  The Fourier transform amplitude exhibits superlattice peaks (orange arrows) indicating the presence of a modulated structure.
  (B), (C) HAADF STEM images at 293K and 93K, respectively. 
  (D), (E) Mapping of transverse, commensurate periodic lattice displacements at 293K and 93K, respectively. 
  Blue (yellow) arrows correspond to cation displacements oriented 90$^{\circ}$(-90$^{\circ}$) relative to $\mathbf{q}$.
  Area of arrows scales linearly with the magnitude of displacements.}
  \label{F:Fig2}
\end{figure}

To overcome challenges associated with intensity-based techniques, we characterize $\mathbf{\Delta(r)}$ using phase-sensitive, real space STEM.
Figure~\ref{F:Fig2}A shows a high-angle annular dark-field (HAADF) STEM lattice image at 93K; Bi/Sr/Ca columns (green) appear bright, Mn columns (red) appear faint, and O atoms are invisible. 
The contrast is due to the dependence of the scattering cross-section in HAADF-STEM on the atomic number.
The temperature of the sample ($\sim$93K) is directly read from a thermocouple near the tip of the sample rod, but the true temperature may be slightly higher.
To minimize stage drift and noise, stacks of 40 fast-acquisition (0.5$\mu$s/pixel) images are collected, registered by cross-correlation, and averaged.
The data demonstrate STEM imaging near liquid-nitrogen temperature ($\sim$ 93K) with high resolution ($\sim$ 0.78\r{A}) and signal-to-noise ratio (\textit{SI Appendix}, Fig. S2), which may be combined with picometer precision mapping of atomic columns.

The FT (\textit{inset}) of the lattice image exhibits sharp superlattice spots (\textit{arrows}), allowing mapping of lattice modulations associated with superlattice peaks\cite{Savitzky2017}.
We damp the amplitude of the superlattice peaks to the background level and apply an inverse Fourier transform.
The result is a reference lattice image in which the targeted modulation has been removed (\textit{SI Appendix}, Fig. S3).
By fitting atomic columns using two-dimensional Gaussian functions in both the original lattice and the reference lattice, we may obtain lattice shifts associated with the charge-ordered state.  
As previously discussed, the method accurately yields the structure of the modulation except at atomically sharp discontinuities in the modulation field\cite{Savitzky2017}.

Figures~\ref{F:Fig2}B and~\ref{F:Fig2}D show a STEM image at 93K and the corresponding mapping of the lattice response, respectively. 
The arrows correspond to displacements of atomic columns in the original image relative to the generated reference lattice, and the color represents the angle of the displacement vector relative to $\mathbf{q}$, with blue (yellow) corresponding to 90$^{\circ}$(-90$^{\circ}$). 
Thus, the low temperature ordered state in BSCMO involves transverse, displacive modulations of both the Bi/Sr/Ca sites and the Mn sites, with amplitudes in the range of $6-11$ pm.
For comparison, Figs.~\ref{F:Fig2}C and~\ref{F:Fig2}E show a HAADF image and the corresponding PLD mapping at 293K.
In these well-ordered regions, lattice displacements appear commensurate with the lattice and exhibit $3a$ periodicity at both temperatures.
In contrast, area-averaged diffraction suggests an average $3.18a$ ($3.01a$) periodicity at 293K (93K), which exposes a discrepancy between local and global measurements of the modulation wavevector.

\begin{figure*}
  \includegraphics[width=\textwidth]{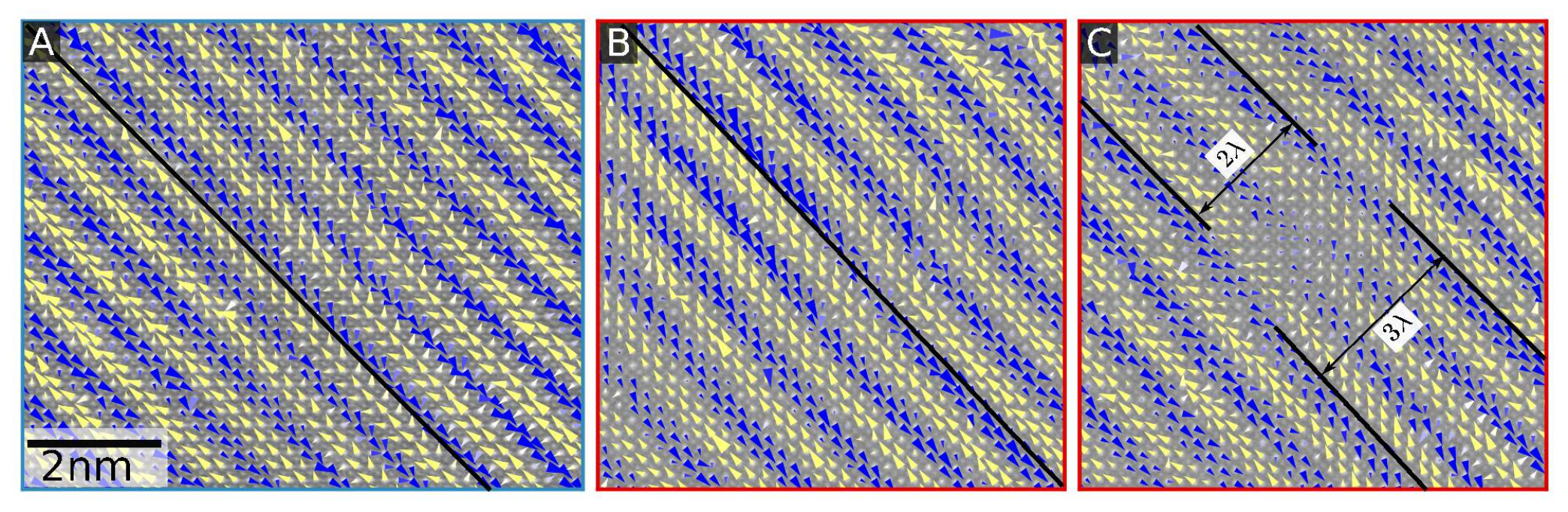}
  \caption{Local variations and disorder of stripes. (A), (B) Shear deformation of striped modulations at 93K and 293K, respectively. 
  A shear deformation appears as a bending of the wavefronts. 
  The black line traces the direction perpendicular to the wavevector and helps visualize the deformation of the wavefront. 
  (C) Stripe dislocation at 293K, in which one wavefront terminates abruptly. 
  }
  \label{F:Fig3}
\end{figure*}

Bridging the gap in length scales, we map PLDs over larger areas and show that they undergo spatial variations in their shape and strength (Fig.~\ref{F:Fig3}).
In particular, we observe intrinsic stripe defects including shear deformations, dislocations, and amplitude reduction.
A shear deformation is a bending of a wavefront, as shown in Figs.~\ref{F:Fig3}A and~\ref{F:Fig3}B.
Both the 293K map and the 93K map reveal such deformation, with the latter exhibiting a milder and more extended instance.
We also observe other forms of disorder in 293K data including stripe dislocations where a wavefront terminates abruptly (Fig.~\ref{F:Fig3}C).
Displacement magnitudes reduce near defect sites, with dislocations showing a more pronounced reduction.

\begin{figure}[!h]
  \includegraphics[width=\linewidth]{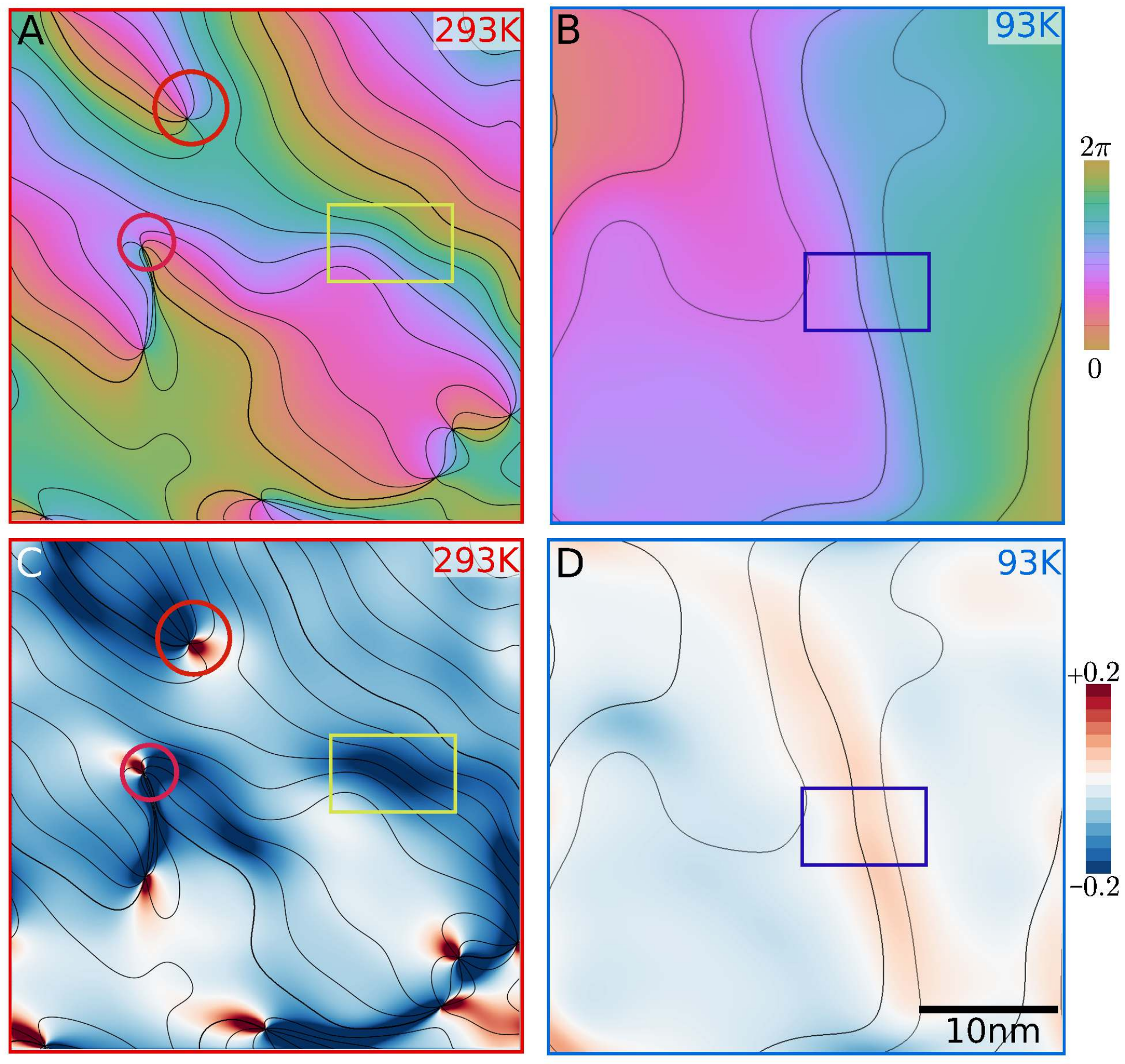}
\caption{Emergent phase coherence at low temperature. (A), (B) Maps of the coarse-grained phase, $\phi(\mathbf{r})$, at 293K and 93K, respectively. 
Black lines represent constant $\pi/4$ phase contours. 
(C), (D) Maps of the phase strain, $\varepsilon_{c}$, at 293K and 93K, respectively. 
Circles correspond to dislocations and boxes correspond to shear deformations. 
} 
  \label{F:Fig4}
\end{figure}

Analysis of the demodulated order parameter, $A(\mathbf{r})^{i\phi(\mathbf{r})}$, unearths nanoscale phase inhomogeneity that coincides with the deformations and dislocations of stripes. 
Phase variations encode deviations from perfect, long-range modulations; they can be extracted using the phase lock-in technique applied previously on spectroscopic scanning tunneling microscopy (STM) data to visualize density wave fluctuations in cuprates\cite{Lawler2010,Mesaros2011} (\textit{SI Appendix, Extracting Coarse-Grained Order Parameter Fields}).
Figure~\ref{F:Fig4}A displays a room temperature phase configuration, $\phi$(\textbf{r}), overlaid with $\pi/4$ constant phase contours (\textit{black lines}).   
The $\phi$(\textbf{r})--map reveals significant spatial inhomogeneity, with $\pm\pi$ phase changes, or more than 4 contours, occurring within regions as small as 5nm.
Even more dramatic variations occur near topological defects (dislocations) where the phase winds by $\pm 2\pi$ around the defect site (circles).
In contrast, we observe in Fig.~\ref{F:Fig4}B a more uniform, slowly varying phase configuration at 93K with a dearth of $\pm 2\pi$ phase change over tens of nanometers.

Phase variations are better visualized via the elastic strain defined by $\varepsilon_{c} = \frac{1}{2}\frac{\mathbf{q_{\perp}}}{q} \cdot \nabla \phi(\mathbf{r})$\cite{Feinberg1988}. 
Strain is normalized such that a $\pm$2$\pi$ phase shift over one wavelength ($\lambda$=2$\pi/q$) corresponds to $\pm$1, and a positive (negative) phase strain represents a local compression (expansion) of the wavefronts (\textit{SI Appendix}, Fig. S5). 
In Figs.~\ref{F:Fig4}B and~\ref{F:Fig4}C, we show $\varepsilon_{c}$ maps overlaid with phase contours at 293K and 93K, respectively. 
We observe at room temperature a moderate background strain, punctured by regions of large, localized phase gradients, notably near dislocations (circle) and shear deformations (rectangle).
The $\varepsilon_{c}$--map at 93K exhibits mild variations and much smaller strain values throughout the full field of view, an indication of emergent phase homogeneity well below T$_{c}$.

Gradients in the phase result in superlattice peak shifts in the FT amplitude, assuming variations do not average to zero (\textit{SI Appendix, Phase Variations and Wavevector Shifts}, Figs. S5 and S6).
The shift is given by $\delta\mathbf{q} = \mathbf{q-q_{0}}=\langle\nabla \phi(\mathbf{r})\rangle$, where $\mathbf{q_{0}}$ is the lattice-locked, commensurate wavevector and $q_{0}=1/3$ r.l.u in our case. 
Therefore, the reduction in phase gradients at 93K, i.e. $\langle\nabla \phi(\mathbf{r})\rangle \approx 0$, is consistent with $q$ approaching $q_{0}$ in diffraction measurements (Fig.~\ref{F:Fig1}D).
As a corollary, negative phase strain at room temperature results in a reduction of the average $q$ measured in diffraction.
Previous work on incommensurate order invokes models based on domain walls, competition with ferromagnetism, or Fermi surface effects to explain the changes in wavevector\cite{Chen1999,Milward2005}.
Our data advance instead that the modulations remain locally locked to the underlying lattice, with wavevector changes reflecting deformations of the phase-field, and that incommensurate-commensurate transitions are an indication of emergent phase homogeneity at low temperature.
We reiterate that, due to phase variations, intensity-based probes will detect wavevector shifts, which are consistent with several distinct local structures. 
On the other hand, phase sensitive probes including STEM can measure the underlying periodicity at the atomic scale.

\begin{figure}[!h]
  \includegraphics[width=.8\linewidth]{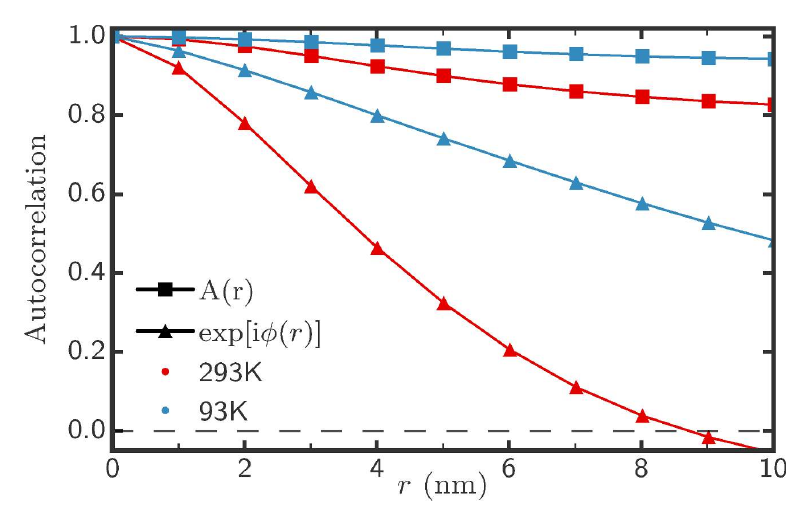}
\caption{Autocorrelations of the phase component and the amplitude component at 293K (red) and 93K (blue). 
  Lines are guides for the eye.
  The slow decay of amplitude correlations at both temperatures suggests that the amplitude is not the main driver of long range order.
  In contrast, the strong temperature-dependent decay of phase correlations supports that phase variations govern long range order. } 
  \label{F:Fig5}
\end{figure}

Having revealed temperature-dependent phase disorder, we address the relative weights of amplitude and phase variations by calculating their respective autocorrelations.
In Fig.~\ref{F:Fig5}, we observe that amplitude correlations quickly plateau to $\sim 0.9$ at 293K and $\sim 0.8$ at 93K over accessible length scales ($\sim 10 nm$ for reasonable statistics); their slow decay at both temperatures suggests they do not influence the correlation length of stripes.  
Phase correlations, on the other hand, decay rapidly at 293K, becoming completely uncorrelated beyond $\sim 8nm$.
At 93K, they remain finite and relatively significant ($\sim 0.5$) beyond $10nm$. 
The strong temperature dependence of phase correlations suggests that the phase component is the primary driver of long range order. 
The situation invites an analogy to disordered superconductors where the loss of phase coherence can cause a superconductor-to-insulator transition despite the persistence of a finite amplitude\cite{Emery1995}. 

Finally, local inspection of amplitude fields at 293K and 93K (\textit{SI Appendix}, Fig. S4) reveals an interaction between phase variation and amplitude variation.
From a mean field perspective, amplitude fluctuations should be suppressed because they cost finite energy.
When the phase varies slowly, the amplitude is expectedly robust and uniform.
However, in regions of large phase gradients, the amplitude weakens, particularly at dislocation sites where it collapses completely (\textit{SI Appendix, Phase Variation and Amplitude Variation}, Fig. S4).
Phase deformations increase the elastic energy density by an amount proportional to $|\nabla \phi(\mathbf{r})|^{2}$, which may require amplitude suppression in regions of diverging phase gradients\cite{Hall1988,coppersmith1991diverging}.
Our observations are consistent with this picture, and attest to the importance of phase fluctuations in shaping the strength and correlations of charge order.

\subsection*{Discussion}

Based on these results, we propose that experimental observations of coincident incommensuration and emergent competing states are related to the local disordering of the phase component; 
phase gradients account for (i) the reduction of the correlation length, (ii) apparent changes in the average wavevector, and (iii) the local quenching of the amplitude which may allow another order to materialize.
This may explain the competition between ferromagnetism and incommensurate charge order in manganites\cite{Chen1996,Milward2005}, or even the granular interplay between short range incommensurate charge density waves and superconductivity in cuprates or transition metal dichalcogenides. 

A successful microscopic theory for charge order, and the many exotic states it affects, should account for all relevant degrees of freedom and for the possibility of inhomogeneity.
Using cryogenic STEM on a manganite, we have directly measured the lattice component and found that the cations undergo transverse displacements relative to the modulation wavevector.
Models based solely on separation of charge or orbital ordering are therefore insufficient.
Atomic displacements change bond distances, bond angles, and hence exchange interactions, and may be key to explaining how charge order impacts other electronic states.   
We have also found that the underlying ground state is locally commensurate with the lattice at both room and cryogenic temperatures despite an apparent incommensuration in reciprocal space, reinforcing both the importance of lattice locking and the fundamental effects of nanoscale phase inhomogeneity on macroscopic behavior. 
We envision that mapping the lattice component using cryogenic STEM in correlated materials will elucidate structural ground states and reveal connections between atomic displacements and various symmetries of electronic and orbital order.

\clearpage

\clearpage

\section*{Acknowledgements}
Research was primarily supported by the Department of Defense, Air Force Office of Scientific Research under award number FA 9550-16-1-0305. 
We also acknowledge support by the Packard Foundation. 
This work made use of the Cornell Center for Materials Research Shared Facilities which are supported through the NSF MRSEC program (DMR-1120296). 
The FEI Titan Themis 300 was acquired through NSF-MRI-1429155, with additional support from Cornell University, the Weill Institute and the Kavli Institute at Cornell.
B.H.S. was supported by NSF GRFP grant DGE-1144153. 
The work at Rutgers was supported by the Gordon and Betty Moore Foundation’s EPiQS Initiative through Grant GBMF4413 to the Rutgers Center for Emergent Materials.

\section*{Author Contributions}
I.E.B., R.H., and L.F.K. designed and performed research; I.E.B, B.H.S., R.H., and L.F.K. analyzed data; A.S.A, J.K. and S.W.C. grew single crystals and performed resistivity measurements; and I.E.B and L.F.K. wrote the paper.

\section*{Competing Interests Statement}
The authors declare that they have no competing financial interests.

\section*{Materials \& Correspondence} 
Corresponsence and requests for materials should be addressed to L.F.K. \newline (lena.f.kourkoutis@cornell.edu).

\bibliography{20170729_bib_edited_v2}

\end{document}


\title{Supplemental Information -- Commensurate Stripes and Phase Coherence in Manganites Revealed with Cryogenic Scanning Transmission Electron Microscopy}
\author{Ismail El Baggari} 
\affiliation{Department of Physics, Cornell University, Ithaca, NY 14853, USA}%
\author{Benjamin H. Savitzky} 
\affiliation{Department of Physics, Cornell University, Ithaca, NY 14853, USA}%
\author{Alemayehu S. Admasu} 
\affiliation{Rutgers Center for Emergent Materials and Department of Physics and Astronomy, Rutgers University, Piscataway, NJ 08854, USA}%
\author{Jaewook Kim} 
\affiliation{Rutgers Center for Emergent Materials and Department of Physics and Astronomy, Rutgers University, Piscataway, NJ 08854, USA}%
\author{Sang-Wook Cheong} 
\affiliation{Rutgers Center for Emergent Materials and Department of Physics and Astronomy, Rutgers University, Piscataway, NJ 08854, USA}%
\author{Robert Hovden} 
\affiliation{School of Applied and Engineering Physics, Cornell University, Ithaca, NY 14853, USA}%
\affiliation{Current address: Department of Materials Science and Engineering, University of Michigan, Ann Arbor, MI 48109, USA}
\author{Lena F. Kourkoutis} 
\affiliation{School of Applied and Engineering Physics, Cornell University, Ithaca, NY 14853, USA}%
\affiliation{Kavli Institute for Nanoscale Science, Cornell University, Ithaca, NY 14853, USA}

\date{\today}

\maketitle

\section{Materials and Methods}
\begin{figure}
  \includegraphics[width=.8\linewidth]{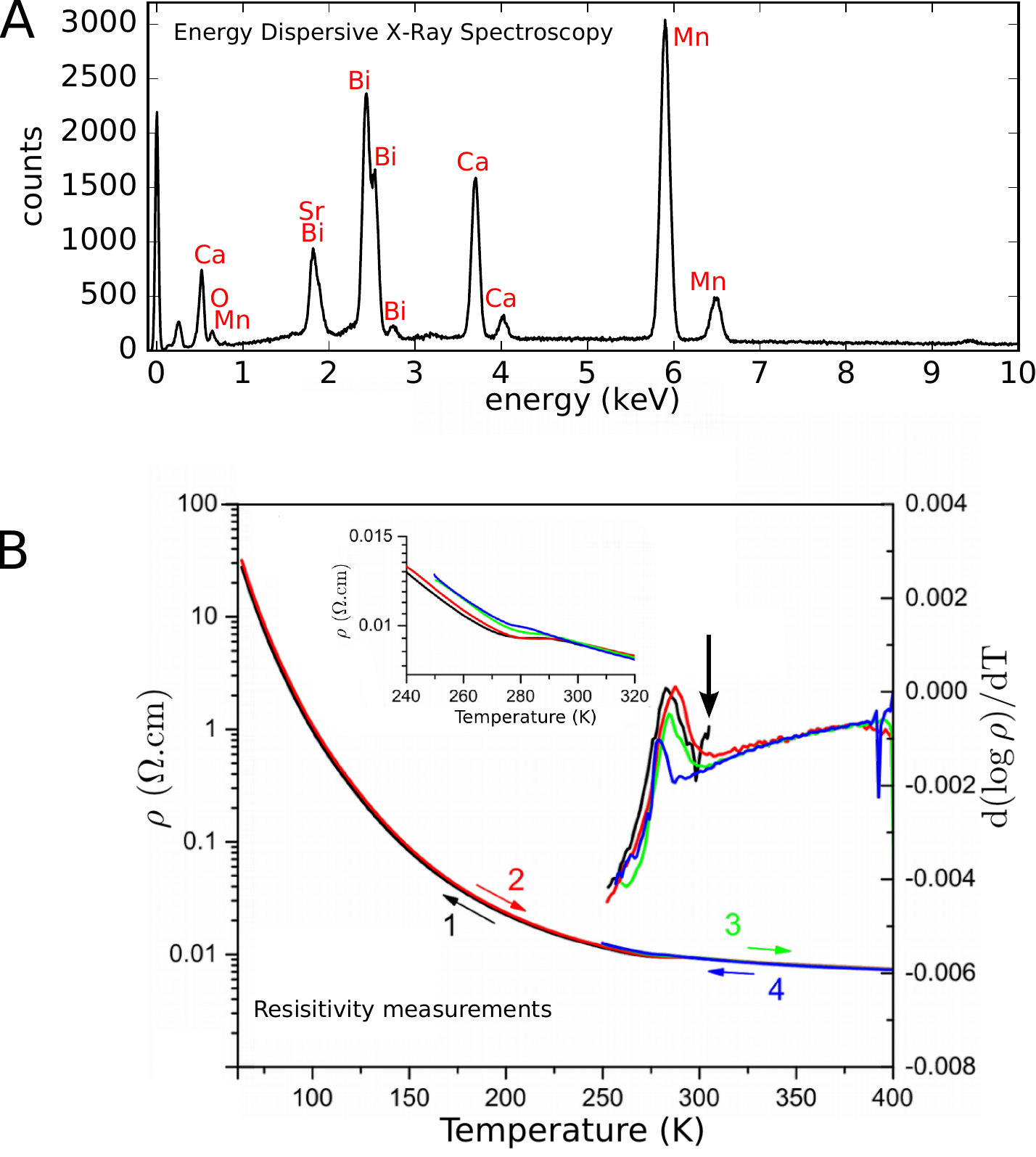}
  \caption{Composition and transport properties of BSCMO. (A) Using energy dispersive X-ray spectroscopy on Bi$_{1-x}$Sr$_{x-y}$Ca$_{y}$MnO$_{3}$, the sample composition is approximately $x=0.65$ and $y=0.47$
           (B) Transport measurements as a function of temperature.
           In the inset, we see a change in the slope of the resistivity curve near T$_{c}$ $\sim$ 300K, which is associated with a charge ordering transition. 
           The transition is broad and gradual and is best visualized in the derivative of the logarithm of the resistivity ($\textrm{black arrow}$).
           The resistivity curve has a slight thermal cycling dependence.
           The arrows and colors indicate heating (red, green) or cooling (black,blue).
}
  \label{F:edxTransport}
\end{figure}

Bi$_{1-x}$Sr$_{x-y}$Ca$_{y}$MnO$_{3}$ (BSCMO) single crystals are grown using the flux method, using Bi$_2$O$_3$, CaCO$_3$, SrCO$_3$, and Mn$_2$O$_3$, as reported previously\cite{Savitzky2017}.
Sample preparation for electron microscopy and energy dispersive X-ray spectroscopy (EDX) are performed on a FEI Strata 400 Focused Ion Beam (FIB). 
From EDX, the composition is determined to be approximately $x=0.65$ and $y=0.47$ (Fig.~\ref{F:edxTransport}A), with negligible variations over the whole sample (size 0.34 $\times$ 0.28 mm). 
Temperature-dependent electrical resistivity measurements show a transition at T$_{c}\approx $ 300K, which is associated with the onset of charge order (Fig.~\ref{F:edxTransport}B).

A thin, electron transparent cross section of BSCMO is extracted using FIB lift out, with estimated thickness in the imaging regions ranging from 10 to 30 nm.
Based on electron diffraction, the orientation of the sample is along the \textbf{b} direction (orthorhombic axis) of the $Pnma$ space group.
At room temperature (293K), BSCMO exhibits satellite peaks, indicating the presence of charge ordering (Fig. 1 \textit{main text}).
We perform electron diffraction and microscopy on an aberration-corrected FEI Titan Themis operating at 300kV. 
Diffraction measurements across temperatures are performed in the same region of the sample using a $\sim 1\mu$m selected area aperture.
In HAADF-STEM, the convergence semi-angle is 30mrad and the collection inner and outer angle are 68mrad and 340mrad, respectively. 
During STEM imaging the sample experiences a $\sim$2 Tesla magnetic field due to its position inside the objective lens, as determined from a Hall bar measurement. 

\section{Cryogenic Scanning Transmission Electron Microscopy}

\begin{figure*}
  \includegraphics[width=.9\textwidth]{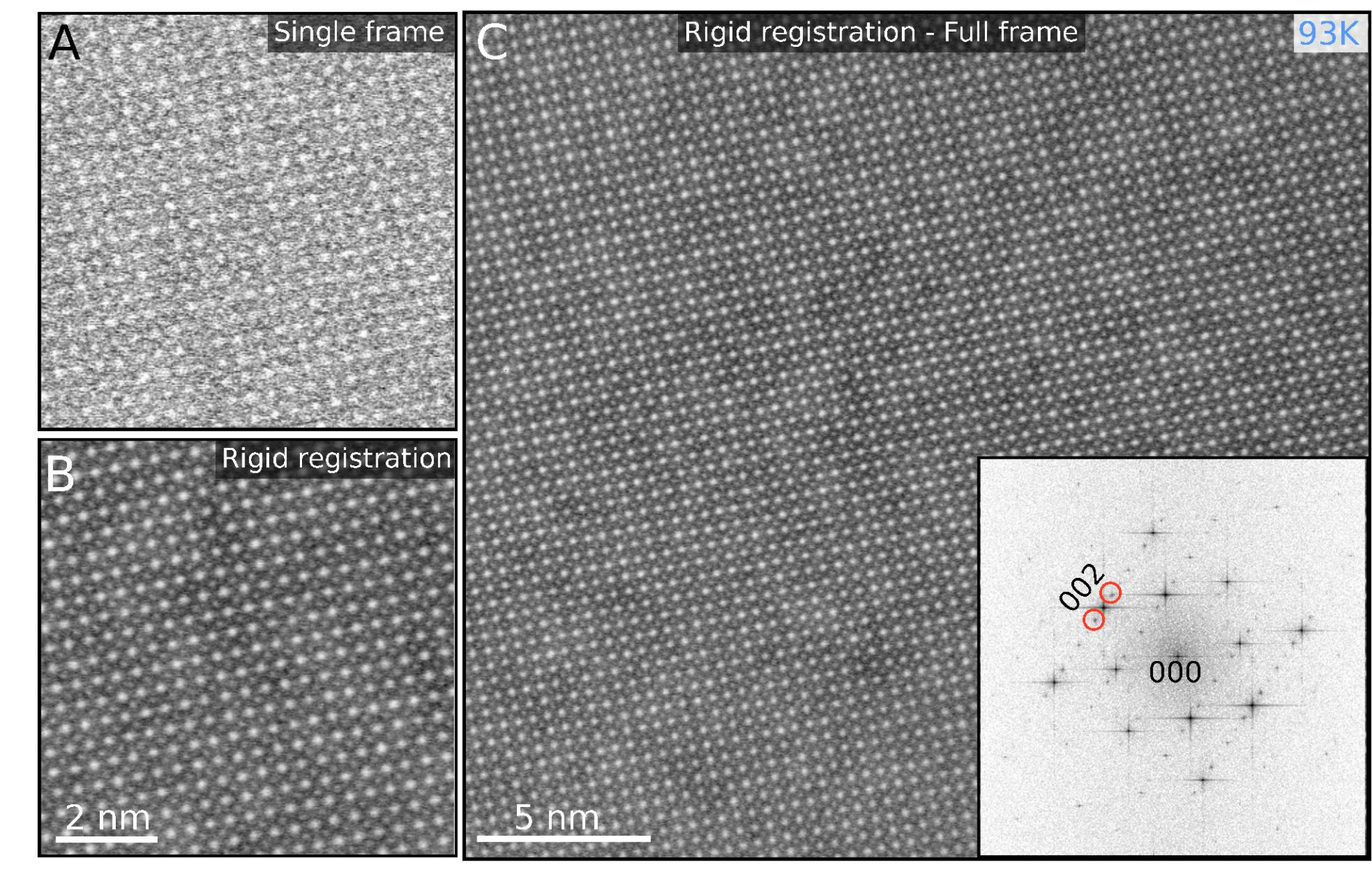}
  \caption{Rigid registration of cryogneic STEM data. (A) A single cryogenic HAADF-STEM image acquired with a 0.5$\mu s$ pixel dwell time. 
  (B) A high signal-to-noise image obtained from averaging 40 rigidly-registered fast-acquisition images. 
  The Bi/Sr/Ca and Mn columns are clearly resolved. 
  Both (A) and (B) are small sections from the full frame image (orange box). 
  (C) The full frame, rigidly-registered HAADF image. 
  Atomic-scale features are well resolved including intensity variations due to cation doping on the A site. 
  (D) The Fourier transform amplitude contains both Bragg peaks and satellite peaks ($inset$, arrows). 
  The information transfer limit estimated from the Fourier transform is $\sim$ 0.78\r{A}.     
 }
  \label{F:registration}
\end{figure*}

For cryogenic experiments, we use a Gatan 636 double-tilt liquid-nitrogen holder. 
The microscope is equipped with a cryogenically cooled box which encloses the sample during low temperature imaging to reduce ice buildup.
After the sample is inserted, we add liquid nitrogen to the holder dewar and wait $\sim$2 hours for the holder to stabilize and for drift to subside.
The temperature is read from a thermocouple near the tip of the sample rod, but the true sample temperature may be slightly different. 
Due to reduced stability at cryogenic temperatures, stacks of 20 to 40 fast-acquisition (0.5$\mu s$/pixel dwell time) images are collected, registered by cross-correlation, and averaged to minimize stage drift and noise.
Acquisition parameters are optimized for Fourier space sampling, field-of-view, pixel density, signal-to-noise ratio, and minimal image distortion. 
The data in Fig. 2 of the main text is sampled at 17.6pm/pixel and the data in Figs. 3 and 4 is sampled at 38.5pm/pixel. 
While the signal-to-noise ratio in one fast-acquisition image is low (Fig.~\ref{F:registration}A), the rigidly-registered and averaged image in Figs.~\ref{F:registration}B and~\ref{F:registration}C show that atomic columns are well resolved and that the signal-to-noise ratio is high.
The information transfer estimated from the highest frequency peak in the Fourier transform (FT) is $\sim$0.78\r{A} (Fig.~\ref{F:registration}D).

\section{Periodic Lattice Displacement Mapping}
\begin{figure*}
  \includegraphics[width=.8\textwidth]{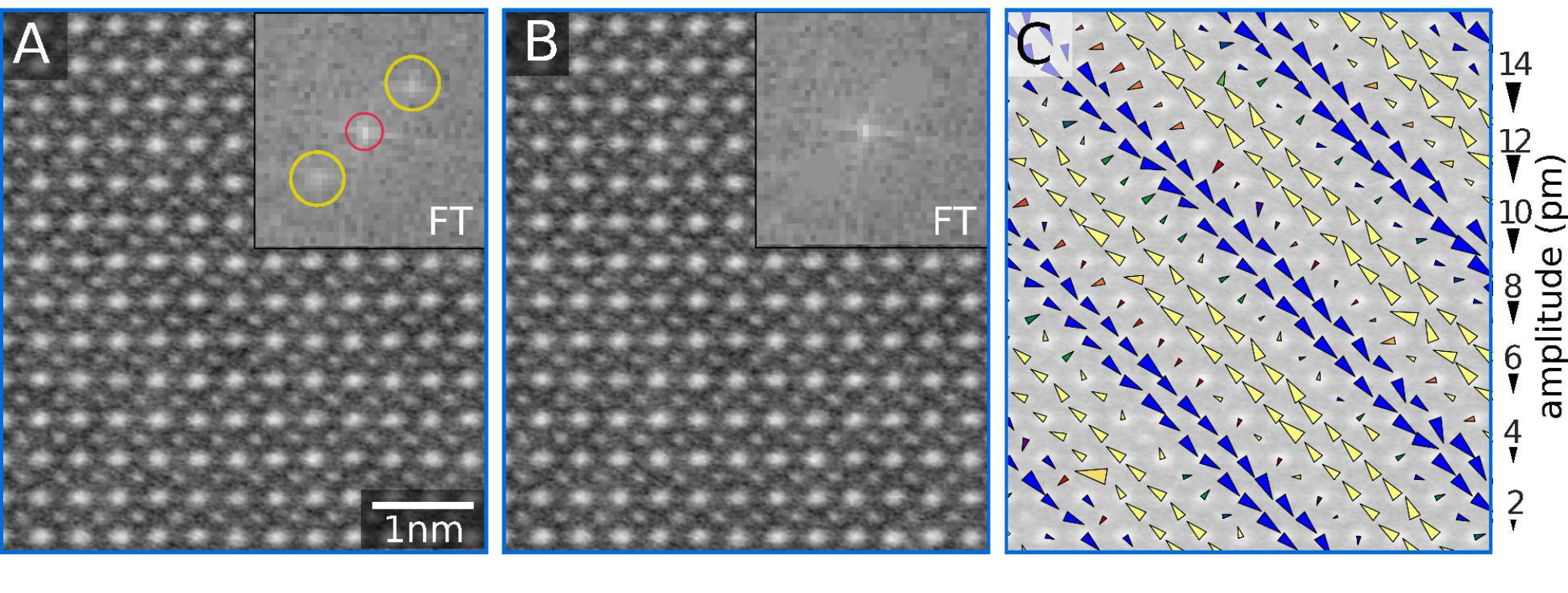}
  \caption{Mapping transverse periodic lattice displacements in BSCMO. (A) Cryogenic HAADF STEM image (cropped). 
The Fourier transform amplitude exhibits Bragg peaks and satellite peaks (see main text). 
The inset shows an example of a Bragg peak (red circle) accompanied by satellite peaks (yellow circles). 
(B)  Reference lattice image generated by damping all satellite peaks corresponding to the modulation to the background level. 
The inset shows a section of the FT with damped satellite peaks. 
(C) Mapping of displacements associated with satellite peaks. 
The displacements are obtained by fitting 2D Gaussians to atomic columns in original HAADF image and the reference image.    
 }
  \label{F:method}
\end{figure*}

The method for mapping periodic lattice displacements in STEM data is described in great detail elsewhere\cite{Savitzky2017}.
Briefly, the FT amplitude shows satellite peaks decorating lattice Bragg peaks. 
The superlattice peaks, which correspond to a modulation with wavevector $\mathbf{q}$, are sharp and decoupled in Fourier space (Fig.~\ref{F:method}A). 
We identify all satellite peaks and damp their amplitude to the background level while maintaining phase information (Fig.~\ref{F:method}B).
Background levels are determined using two-dimensional Gaussian fits to satellite peaks.
By applying an inverse Fourier transform of the processed FT, we obtain a lattice image where the $\mathbf{q}$-modulation has been removed (Fig.~\ref{F:method}B).
We note that the amplitude damping must encompass the satellite peak completely in order to capture all information about the modulation and that, once fully captured, the mapping is insensitive to increasing mask size.
We may extract atomic column positions and measure displacements of atomic columns in the original data (Fig.~\ref{F:method}A) relative to the reference data (Fig.~\ref{F:method}B), as shown in Fig.~\ref{F:method}C.
In BSCMO, we find cation displacements which are transverse and with maximal amplitudes on the order of $6-8pm$.

\section{Extracting Coarse-Grained Order Parameter Fields}

As described previously for scanning tunneling microscopy data\cite{Lawler2010}, we extract a coarse-grained phase field, $\phi(\mathbf{r})$, associated with the $\mathbf{q}$-modulation.
We first Fourier filter regions surrounding a superlattice peak, using a Gaussian filter with a width $\sigma = L^{-1}$ where $L$ is a coarsening length-scale which we choose to be on the order of two wavelengths.
The resultant real space image has all periodicities removed, except for the one associated with the $\mathbf{q}$--modulation.
The filtered image may be described by 
\begin{equation*}
\tilde{I}(\mathbf{r}) \sim \sin(\mathbf{q}\cdot\mathbf{r}+\phi(\mathbf{r}))
\end{equation*}
The filtered image, however, does not contain perfect modulations due to order parameter (OP) inhomogeneity. 
To extract the phase field responsible for said inhomogeneity, we apply the phase lock-in technique\cite{Lawler2010}:

\begin{itemize}
\item We generate two reference signals $\sin(\mathbf{q}\cdot\mathbf{r})$ and $\cos(\mathbf{q}\cdot\mathbf{r})$ 
\item We multiply the filtered image by the two reference signals
\item We obtain $X(\mathbf{r})$ and $Y(\mathbf{r})$ where
\begin{equation*}
\begin{aligned}
\begin{cases}
X(\mathbf{r}) = \sin(\mathbf{q}\cdot\mathbf{r}) \sin(\mathbf{q}\cdot\mathbf{r}+\phi(\mathbf{r})) \\
Y(\mathbf{r}) = \cos(\mathbf{q}\cdot\mathbf{r}) \sin(\mathbf{q}\cdot\mathbf{r}+\phi(\mathbf{r}))
\end{cases}
\end{aligned}
\end{equation*}
\begin{equation*}
\begin{aligned}
\begin{cases}
X(\mathbf{r}) = \frac{1}{2} (\cos\phi(\mathbf{r}) - \cos(2\mathbf{q}\cdot\mathbf{r}+\phi(\mathbf{r})) ) \\
Y(\mathbf{r}) = \frac{1}{2} (\sin\phi(\mathbf{r}) + \sin(2\mathbf{q}\cdot\mathbf{r}+\phi(\mathbf{r})) )
\end{cases}
\end{aligned}
\end{equation*}
\item We low pass filter $X(\mathbf{r})$ and $Y(\mathbf{r})$ to remove of the second, high-frequency terms obtaining 
\begin{equation*}
\begin{aligned}
\begin{cases}
\tilde{X}(\mathbf{r}) \approx \cos\phi(\mathbf{r})  \\
\tilde{Y}(\mathbf{r}) \approx \sin\phi(\mathbf{r}) 
\end{cases}
\end{aligned}
\end{equation*}
\item The coarse grained phase is given by 
\begin{equation*}
\phi(\mathbf{r}) = \arctan[\tilde{Y}(\mathbf{r})/ \tilde{X}(\mathbf{r})]
\end{equation*}
\end{itemize}

\section{Phase variation and amplitude variation}
In Fig. 3 of the main text, we observe that the magnitude  of $\mathbf{\Delta(r)}$ weakens in regions of shear deformations and dislocations.
By separating and extracting the phase and amplitude components, we reveal that large phase gradients are accompanied by amplitude reduction in both room and cryogenic temperature data (Figs.~\ref{F:correspondence}C and~\ref{F:correspondence}D).
In particular, near shear deformations (rectangles) and dislocations (circles), the value of the square of the phase gradient increases (Figs.~\ref{F:correspondence}A and~\ref{F:correspondence}B), which indicates that these defects are driven by phase variations, and the amplitude weakens, which indicates that there is a coupling between the two components.
When the phase varies slowly, the amplitude component is strong and uniform.
As shown in Fig. 5 in the main text, amplitude correlations decay negligibly, indicating that they do not influence long range order.
Nevertheless, their modest temperature dependence may be attributed to their coupling to large phase gradients.

We propose a simple Ginzburg-Landau (GL) theory that captures the interplay between phase disorder and amplitude variation.
Consider the order parameter given by $\psi(\mathbf{r}) = A(\mathbf{r})\textrm{e}^{i\phi(\mathbf{r})}$, where, for simplicity, we neglected the vectorial nature of the amplitude component. 
The GL energy density, $f$, is given by\cite{mcmillan1975landau,Lee1979,Hall1988} 
\begin{equation*}
f  =  a|\psi(\mathbf{r})|^{4} + b|\psi(\mathbf{r})|^{2} + c|\nabla \psi(\mathbf{r})|^{2} + f_{\textrm{imp}}
 \end{equation*}
The first two terms are the customary terms in a GL expansion. 
The $a$, $b$ and $c$ are phenomenological constants with $a>0$, $c>0$, and $b=(T-T_{c})/T_{c} < 0$ for $T< T_{c}$.
The third term represents the energy due to order parameter variations, and the fourth term represents the energy due to coupling to impurities.
We assume that an impurity at site $\mathbf{R_{i}}$ fixes the value of the phase component to $\Theta(\mathbf{R_{i}})$.
There may be other sources for phase distortion including boundary conditions, applied fields, and inhomogeneous charge. 
The GL functional becomes
\begin{equation*}
\begin{aligned}
f  = aA^{4}(\mathbf{r}) + bA^{2}(\mathbf{r}) + & c|\nabla A(\mathbf{r})|^{2} +  c{A^{2}(r)} |\nabla \phi(\mathbf{r})|^{2} -\\
& \sum_{i} V_{0} \cos\big(\phi(\mathbf{R_{i}}) - \Theta(\mathbf{R_{i}})\big) 
\end{aligned}
\end{equation*}
where $V_{0}$ controls the coupling strength to the impurity.

The amplitude is expected to vary negligibly due to the quadratic term in the GL energy\cite{Lee1979,Feinberg1988}.
In other words, the amplitude is associated with the condensation energy gain, or gap, due to the formation of the ordered state.
In contrast, phase variations only appear in the gradient term and are more likely to occur.
However, near regions of large phase gradients, the energy density can diverge due to the $|\nabla \phi(\mathbf{r})|^{2}$ term, necessitating changes in the amplitude\cite{coppersmith1991diverging,Hall1988}. 
At a dislocation site, the phase singularity forces amplitude collapse altogether.
Phase variations thus provide a mechanism for the local destruction of the amplitude, which may permit a competing order to emerge. 


\section{Structure Factor of Periodic Lattice Displacements}
A unidirectional modulation breaks the translational and rotational symmetry of the underlying lattice. 
In reciprocal space, the presence of a modulation with wavevector $\mathbf{q}$ is typically evidenced by superlattice peaks at $\pm \mathbf{q}$ near lattice Bragg spots.
The nature of the modulation, whether it is a modulation in the atomic structure factor (charge order, cation order) or a modulation of the atomic positions, affects the Fourier space pattern.
A detailed derivation of the structure factor, S($\mathbf{k}$), of modulated lattices is provided elsewhere\cite{Hovden2016,Savitzky2017}.
Further, the presence of phase disorder distorts the shape and amplitude of superlattice peaks.

Here, we briefly derive S($\mathbf{k}$) for a lattice in which the atomic sites, $\{\mathbf{R}\}$, undergo periodic lattice displacements, $\mathbf{\Delta(\mathbf{r})}=\mathbf{A(r)}\sin\big(\mathbf{q.r}+\phi(\mathbf{r})\big)$, where $\mathbf{q}$ is the wavevector, $\mathbf{A(r)}$ is the amplitude and $\phi(\mathbf{r})$ is the phase.
Lattice positions are given by
\begin{equation*}
\begin{aligned}
\mathbf{R^{'}}&=\mathbf{R} + \mathbf{\Delta(\mathbf{R})}\\
\mathbf{R^{'}}&= \mathbf{R}+ \mathbf{A(R)}\sin\big(\mathbf{q.R}+\phi(\mathbf{R})\big)
\end{aligned}
\end{equation*}
The structure factor is 
\begin{equation*}
\text{S}(\mathbf{k}) = \sum_{\{\mathbf{R^{'}}\}}  \text{exp} \Big[i\mathbf{k}.\mathbf{R^{'}}\Big]  = \sum_{\{\mathbf{R}\}}  \text{exp} \bigg[i\mathbf{k}.\Big({\mathbf{R} + \mathbf{A(R)} \sin \big(\mathbf{q}.\mathbf{R} + \phi({\mathbf{R}}) \big)}\Big)\bigg] 
\end{equation*}
\noindent
Using the identity $\text{exp}[iz\sin(\theta)]=\sum\limits_{\alpha=-\infty}^{+\infty}J_{\alpha}(z) \text{exp}[i\alpha\theta]$ where $J_{\alpha}$ are Bessel functions of the first kind, we get
\begin{equation*}
\begin{aligned}
\text{S}(\mathbf{k}) & = \sum\limits_{\{\mathbf{R}\},\alpha} \text{exp}\big[i\mathbf{k}.\mathbf{R}\big] J_{\alpha}(\mathbf{k.A}) \text{exp}\big[\alpha(\textbf{q.R}+\phi({\mathbf{R}})) \big] \\ 
                     & = \sum\limits_{\{\mathbf{R}\},\alpha} \text{exp}\big[i(\mathbf{k+\alpha q}).\mathbf{R}\big] J_{\alpha}(\mathbf{k.A}) \text{exp}\big[i\alpha\phi({\mathbf{R}})\big]
\end{aligned}
\end{equation*}
The first few terms ($\alpha=0,\pm1, \pm2$) dominate, yielding
\begin{equation}
\begin{aligned}
\text{S}(\mathbf{k}) \approx J_{0}(\mathbf{k.A}) \sum_{\{\mathbf{R}\}}  \mathrm{exp}\big[i\mathbf{k}.\mathbf{R}\big]  + & J_{1}(\mathbf{k.A}) \sum_{\{\mathbf{R}\}}\mathrm{exp}\big[i(\mathbf{k \pm q}).\mathbf{R}\big]  \mathrm{exp}\big[i\phi(\mathbf{R})\big] + \\
& J_{2}(\mathbf{k.A}) \sum_{\{\mathbf{R}\}}\mathrm{exp}\big[i(\mathbf{k \pm 2q}).\mathbf{R}\big]  \mathrm{exp}\big[i2\phi(\mathbf{R})\big]
\end{aligned}
\end{equation}
The first term corresponds to the usual lattice Bragg peaks. The second term corresponds to satellite peaks at $\pm\mathbf{q}$, which are further modulated by a $J_{1}(\mathbf{k.A})$ term and the phase disorder term 
$\mathrm{e}^{i\phi(\mathbf{R})}$.
Higher order harmonics ($\alpha$=2,3...) are also generated by periodic lattice displacements, but their experimental observation may depend on the strength and degree of long range ordering.
For example, in Fig. 1 of the main text, we observe that the second order harmonic peak is barely detectable at room temperature but is well defined at low temperature.
We also point out that the $J_{1}(\mathbf{k.A})$ dependence of satellite peak intensities is indicative of periodic lattice displacements.

\section{Phase Variations and Wavevector Shifts}
In diffraction measurements, the wavevector of a modulation is determined by fitting the Fourier amplitude near the targeted superlattice peak using Lorentzian or Gaussian functions.
Phase variations appear through the area-averaged phase-phase correlation function which affects the shape, width, and amplitude of the Fourier peak.
Here, we argue that the conventional definition of the superlattice peak position does not necessarily represent the true periodicity of a modulation\cite{Mesaros2016};
real space phase gradients are indistinguishable from anisotropic peak shifts in diffraction patterns as well as the STEM FT amplitude.

In the following subsections, we derive the structure factor of modulated lattices in the presence of phase gradients and then calculate the Fourier transforms of simulated, one dimensional modulated lattices.
\subsection{Derivation}
Assuming phase variations are approximately linear over the scattering volume, we have 
\begin{equation*}
\phi(\mathbf{r}) = \phi_{0} + \mathbf{\nabla}\phi(\mathbf{r})\cdot\mathbf{r}+ ....
\end{equation*}

Plugging into the second term of (1), the satellite peaks are given by
\begin{equation}
\begin{aligned}
\text{S}(\mathbf{k}) & \sim  \sum_{\{\mathbf{R}\}} \delta({\mathbf{r-R}})  \mathrm{exp}\big[i(\mathbf{k \pm q}).\mathbf{r}\big]  \mathrm{exp}\big[i\big(\phi_{0} + \mathbf{\nabla}\phi(\mathbf{r})\cdot\mathbf{r} \big)\big]\\
             & \sim \sum_{\{\mathbf{R}\}} \delta({\mathbf{r-R}})  \mathrm{exp}\Big[i\Big(\mathbf{k \pm q} + \mathbf{\nabla}\phi(\mathbf{r})\Big)\cdot\mathbf{r}\Big]  \mathrm{exp}\big[i\phi_{0}\big]
\end{aligned}
\end{equation}

The linear term causes an anisotropic shift of the wavevector with $\delta{\mathbf{q}}$ = $\nabla \phi(\mathbf{r})$. 
If gradients over the scattering volume average to zero, the net effect is a blurring of the superlattice peak. 
Higher-order terms in the expansion of $\phi(\mathbf{r})$ also contribute to peak broadening.

\subsection{Illustration of the 1D case}
\begin{figure}
  \includegraphics[width=.8\linewidth]{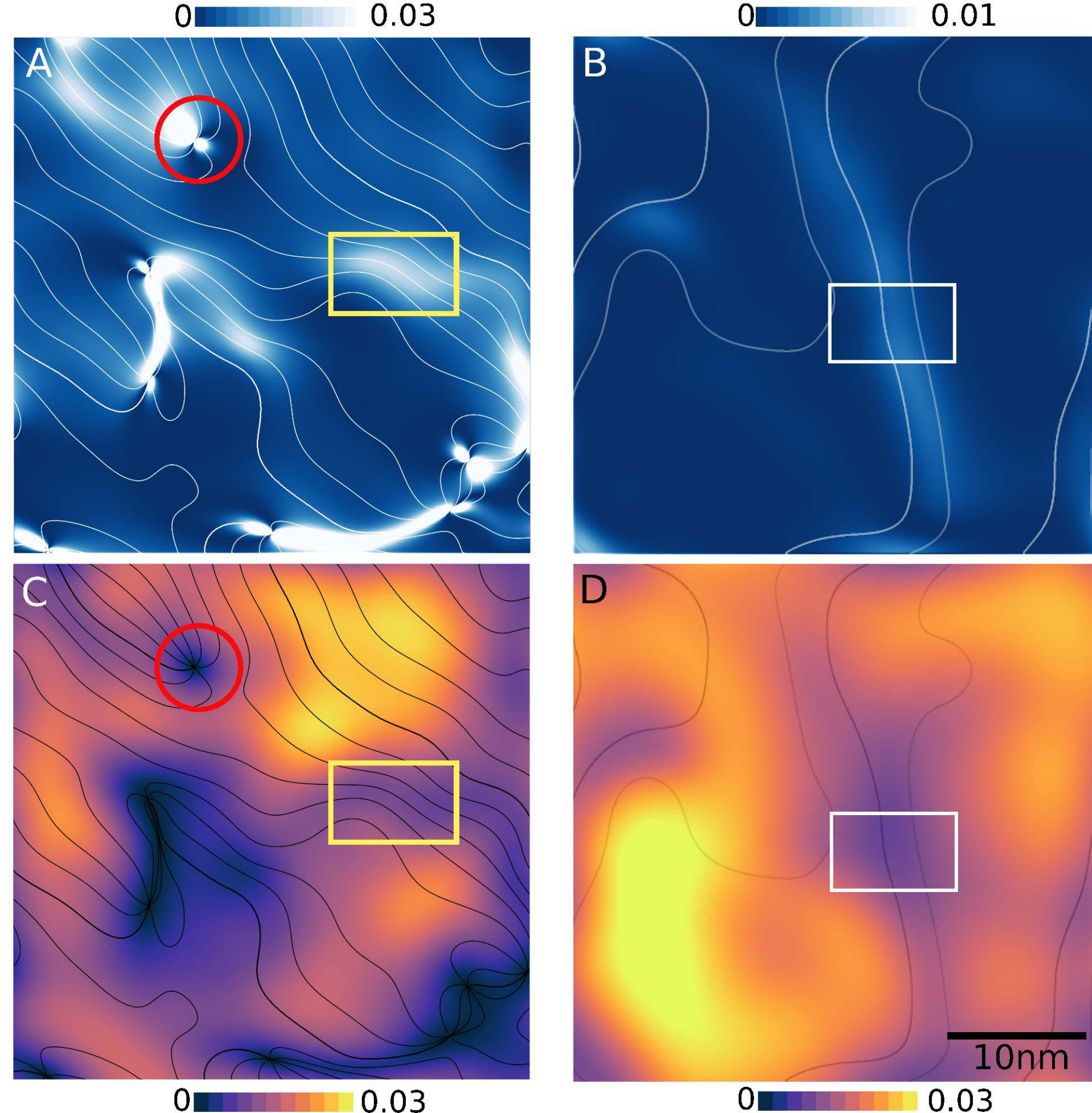}
  \caption{Correspondence between large phase gradients and amplitude reduction. 
  (A), (B) Square of the phase gradient, $|\frac{\mathbf{q_{\perp}}}{q} \cdot \nabla\phi(\mathbf{r})|^{2}$ at 293K and 93K, respectively. 
  (C), (D) Amplitude field, $A(\mathbf{r})$, at 293K and 93K, respectively. 
  The circle corresponds to a dislocation (phase singularity) and the rectangles correspond to shear deformations (phase gradients). 
  We observe amplitude reduction in regions of large phase gradients, suggesting a coupling between phase fluctuations and amplitude fluctuations.
 }
  \label{F:correspondence}
\end{figure}

\begin{figure}
  \includegraphics[width=.6\linewidth]{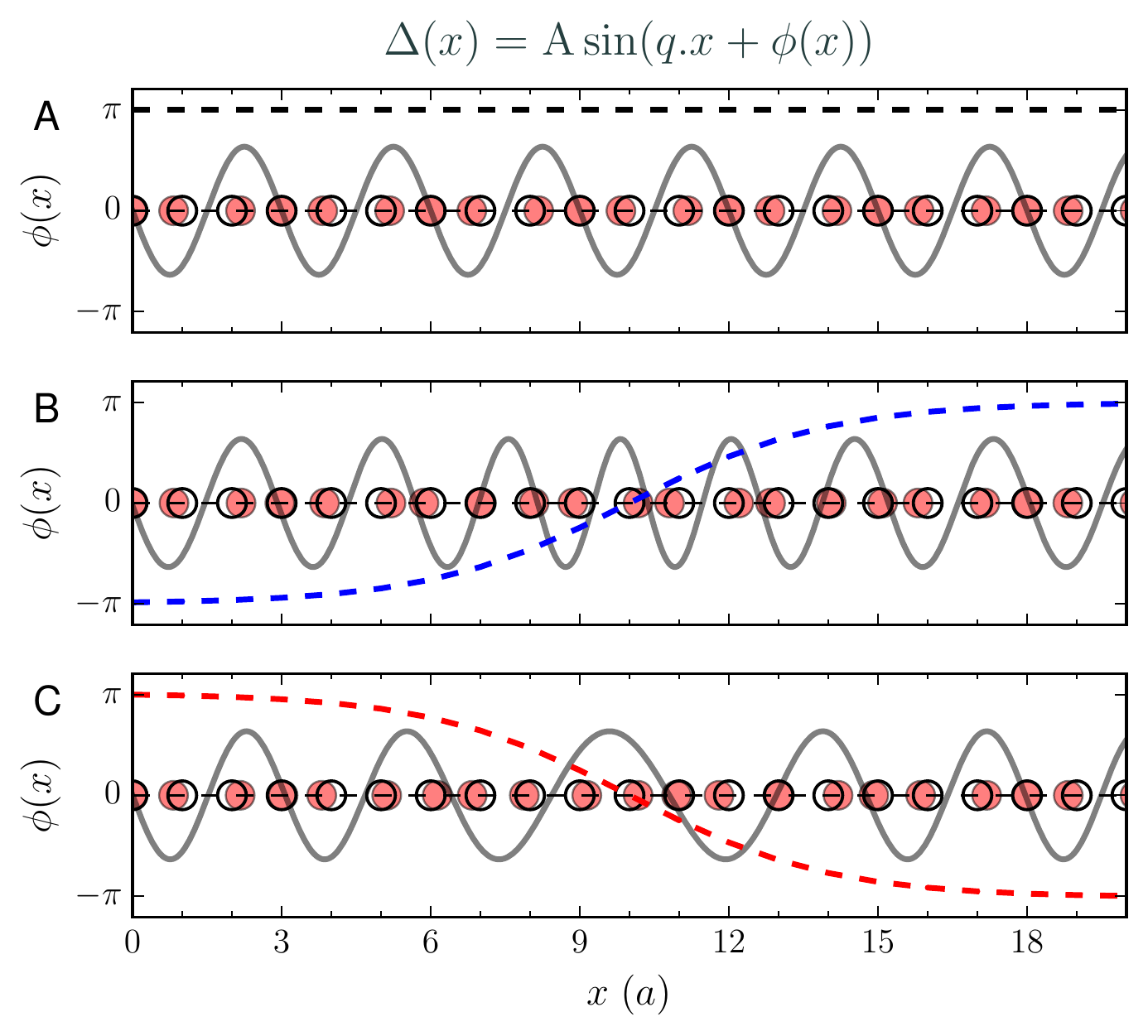}
  \caption{Expansion and compression of wavefronts due to phase gradients. (A)-(C) Phase profiles ($\phi(x)$, dashed lines) and resultant sinusoidal modulations ($\Delta(x)=A\sin(qx+\phi(x))$, gray lines).
  A constant phase (black) corresponds to an ideal sinusoidal modulation. 
  A positive gradient in the phase corresponds to a compression of the wavefront (blue, dashed line). 
  A negative gradient in the phase corresponds to an expansion of the wavefront (red, dashed line). 
  Empty black circles represent the ideal atomic lattice and red circles correspond to the modulated lattice.
  The wavelength of the modulation is $\lambda=3a$.  
 }
  \label{F:phaseProf}
\end{figure}

Consider a one-dimensional displacive modulation given by $\Delta(x)=A\sin\big(qx+\phi(x)\big)$
where $q=1/3 \ (2\pi/a)$ is the wavevector, $A$ is the displacement amplitude, and $\phi(x)$ is the phase.
A constant phase corresponds to an ideal sinusoidal modulation as shown in Fig.~\ref{F:phaseProf}A.
We observe in our data, however, localized regions of significant phase change (Fig.~\ref{F:correspondence}, and Fig. 4 in main text).
A positive (negative) gradient in the phase compresses (expands) wavefronts, which increases (decreases) the number of wavefronts, as shown in Figs.~\ref{F:phaseProf}B and~\ref{F:phaseProf}C.

\begin{figure}
  \includegraphics[width=.5\linewidth]{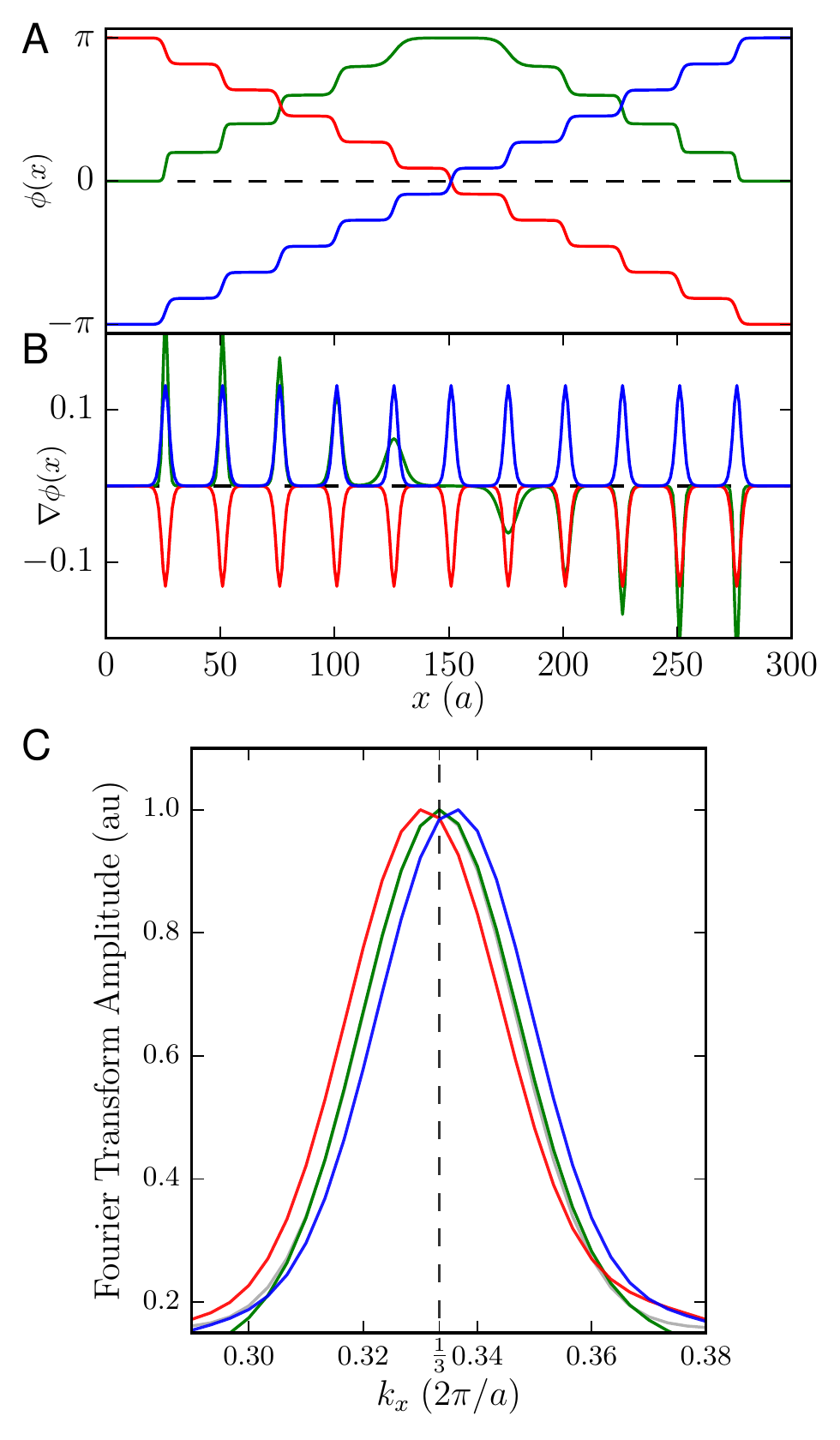}
  \caption{Phase gradients result in wavevector shifts in reciprocal space. (A) 1D phase profiles, $\phi(x)$, with local variations. 
  The black, dashed line represents a constant phase. 
  The red (blue) line represents a step profile with negative (positive) gradients. The green line represents a step profile with gradients averaging to zero.
  The wavelength of the modulation is $\lambda=3a$.  
(B) Gradients of aforementioned phase profiles.
(C) Fourier transform amplitudes of a 1D lattice modulated by $\Delta(x)=\sin(qx+\phi(x))$ with $q=1/3 (2\pi/a)$. 
The color corresponds to phase profiles described before. 
The dashed line, peaked at $q$, corresponds to a constant phase profile.  
 }
  \label{F:phaseFT}
\end{figure}

We now calculate the FT amplitude of simulated modulated lattices in the presence of phase variations.
We generate various phase profiles with smooth, local variations occurring, without loss of generality, every 25 unit cells, and calculate the Fourier transform (FT) amplitudes of the resulting lattices.
The FT amplitudes are convolved with a Gaussian kernel of width $0.01(2\pi/a)$ to simulate the effect of the resolution function.
The black, dashed line in Fig.~\ref{F:phaseFT}A represents a constant phase profile.
As expected, the Fourier transform amplitude (black, dashed line) is peaked at the ordering wavevector $q$, as shown in Fig.~\ref{F:phaseFT}C.
The blue (red) line corresponds to a phase profiles where all variations have positive (negative) phase gradients (Figs.~\ref{F:phaseFT}A and~\ref{F:phaseFT}B).
In Fig.~\ref{F:phaseFT}C, the FT amplitude appears shifted away from $q$, with the blue line moving to a higher $q$ and the red line moving to a lower $q$.
Since a positive (negative) gradient decreases (increases) the wavelength, the wavevector increases (decreases), in agreement with the calculated FT amplitudes.
In the green phase profile, gradients average to zero, yielding a FT amplitude that peaks at $q$ (Fig.~\ref{F:phaseFT}C).

The 1D simulation emphasizes that intensity-based probes do not directly reflect the underlying wavevector of a modulation.
The presence of local phase variations may shift measured Fourier peaks away from the underlying wavevector, supporting that order parameter inhomogeneity alters macroscopic observations of charge order.
However, real-space, phase sensitive probes including STEM or scanning tunneling microscopy allow direct and local measurements of phase inhomogeneity and periodicity of ordered states\cite{Mesaros2016,Mesaros2011}.

\bibliography{20170729_bib_edited_v2}

\clearpage

\begin{figure}
  \includegraphics[width=\linewidth]{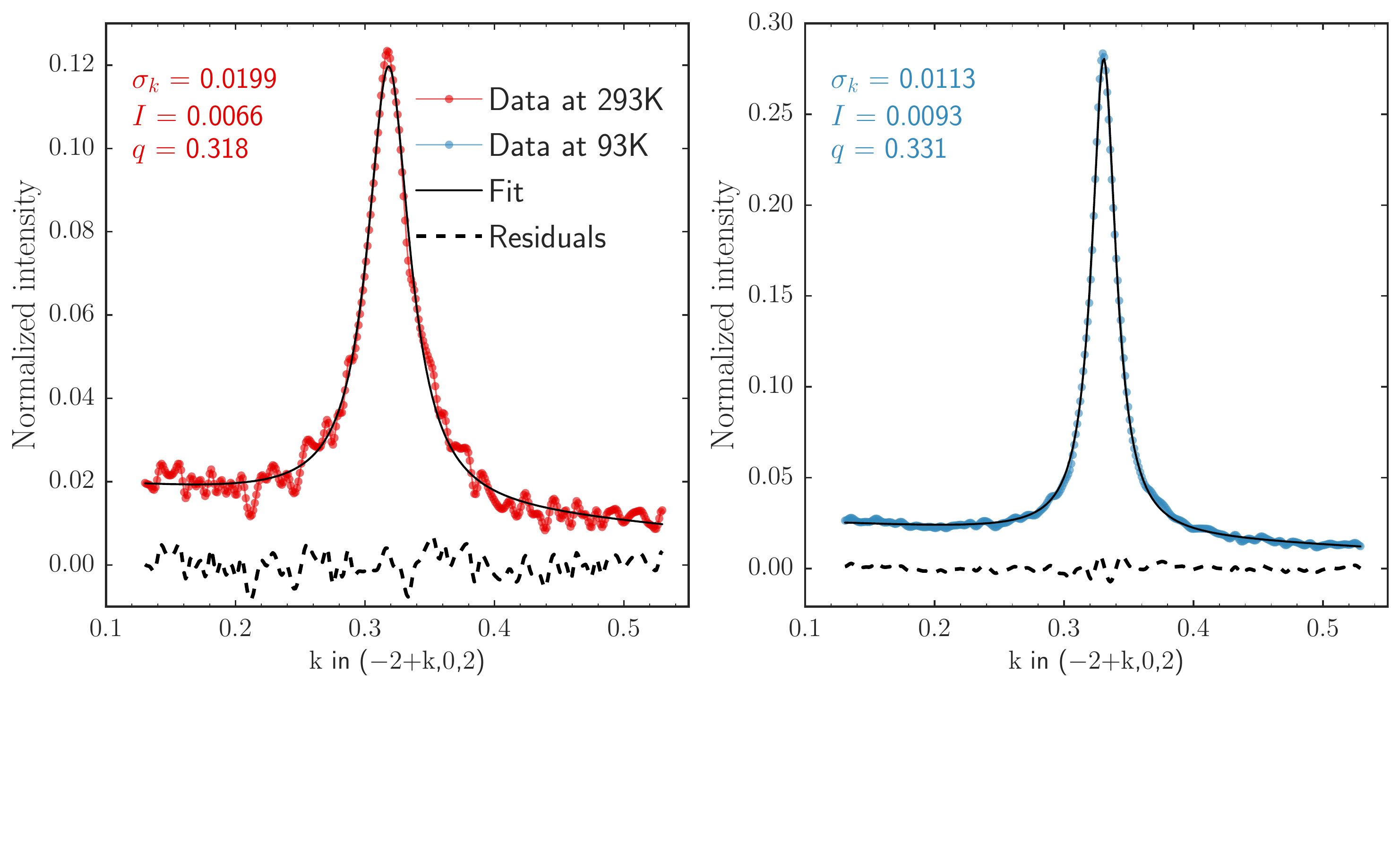}
  \caption{Lorentzian and linear background fit to the superlattice peak profile near the $\bar{2}02$ Bragg peak at 293K ($left$) and 93K ($right$). 
  The Lorentzian function is given by $f(k, I, q,\sigma_{k}) = \frac{I}{\pi}  \frac{\sigma_{k}}{(k-q)^{2} + \sigma_{k}^{2}}$ where $I$ is the intensity (amplitude), $q$ is the wavevector (center), and $\sigma_{k}$ is the width.
  The profile is obtained by integrating between the two ticks in Fig. 1 in the main text and the intensity is normalized by the intensity of the Bragg peak. 
  Upon cooling from 293K to 93K, there is a clear increase in the superlattice peak intensity and a decrease in the full-width-at-half-maximum (2$\sigma_{k}$).
  Note the difference in the scale of the intensity axis.
  We also observe a shift in the superlattice peak position from $q = 0.318$ r.l.u at 293K to $q=0.331$ r.l.u at 93K.
 }
  \label{F:LineCutFit}
\end{figure}

\begin{figure}
  \includegraphics[width=\linewidth]{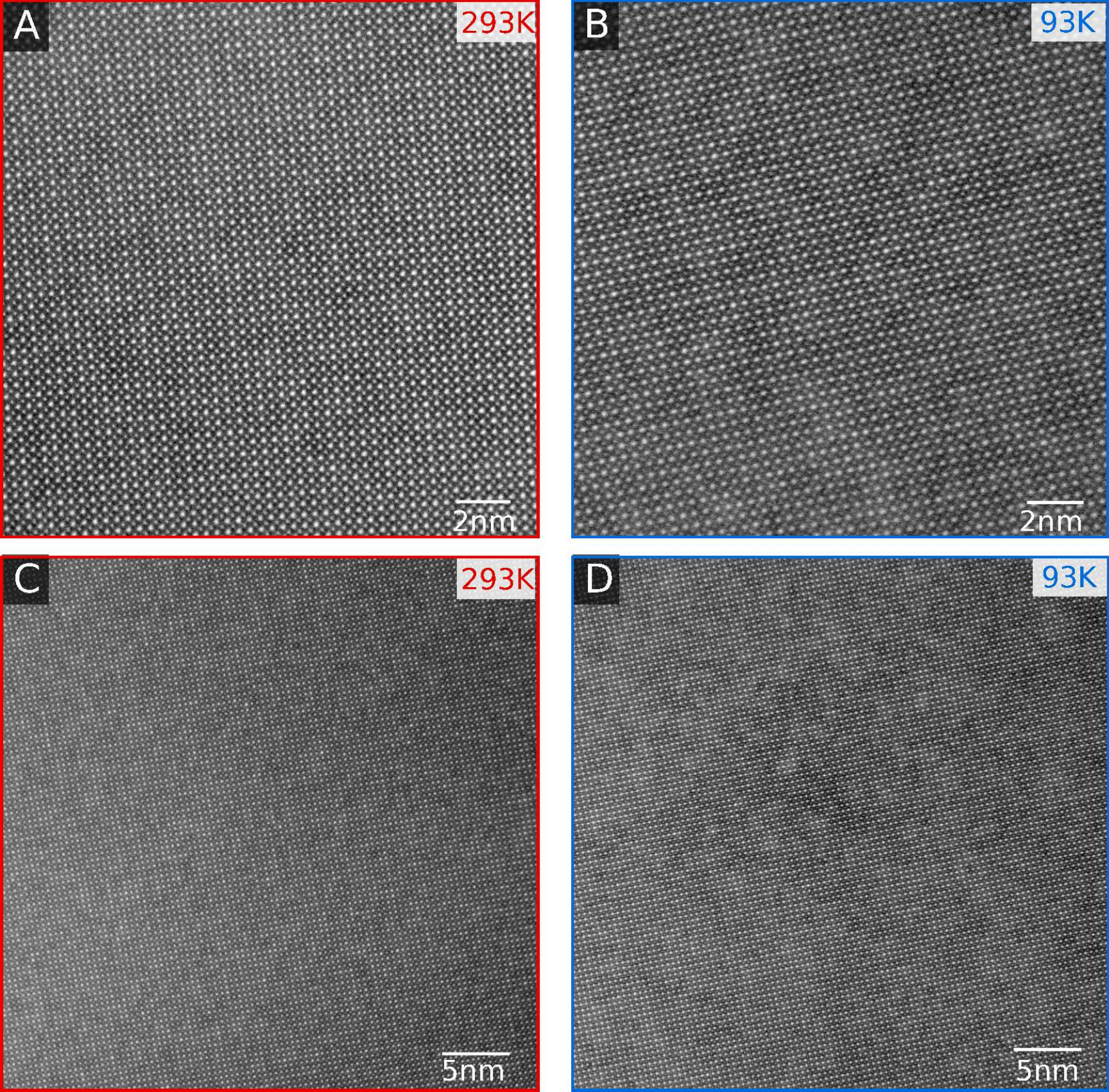}
  \caption{Unprocessed STEM data. (A), (B) Original, full field of view ($\sim$ 20nm) HAADF-STEM data at (A) 293K and (B) 93K corresponding to results in Fig. 2 in the main text.
  		 (A), (B) Original, full field of view ($\sim$ 40nm) HAADF-STEM data at (A) 293K and (B) 93K corresponding to results in Fig.~\ref{F:correspondence} and Figs. 3, 4, 5 in the main text.
         The data are unprocessed except for cross-correlation and a global brightness adjustment.
         Note the varying intensities of atomic columns due to cation doping.
  		  }
  \label{F:data}
\end{figure}